\documentclass[%
preprint,
 amsmath,amssymb,
 aps,
]{revtex4-1}

\usepackage{graphicx}
\usepackage{dcolumn}
\usepackage{bm}

\begin{document}


\title{Structural Self-Assembly and Glassy Dynamics in Locally Adaptive Networks}

\author{Johannes Gr{\"a}wer$^1$, Carl D. Modes$^2$, Marcelo O. Magnasco$^2$, and Eleni Katifori$^1$}
 \affiliation{$^1$Max Planck Institute for Dynamics and Self-Organization (MPIDS), 37077 Goettingen, Germany.}
 \affiliation{$^2$The Rockefeller University, New York, NY 10065}
 \email{katifori@gmail.com}

\date{\today}

\begin{abstract}
Transport networks play a key role across four realms of eukaryotic life: slime molds, fungi, plants, and animals. In addition to the developmental algorithms that build them, many also employ adaptive strategies to respond to stimuli, damage, and other environmental changes. We model these adapting network architectures using a generic dynamical system on weighted graphs and find in simulation that these networks ultimately develop a hierarchical organization of the final weighted architecture accompanied by the formation of a system-spanning backbone. In addition, we find that the long term equilibration dynamics exhibit glassy behavior characterized by long periods of slow changes punctuated by bursts of reorganization events.

\end{abstract}

\pacs{Valid PACS appear here}
\maketitle


\newcommand{\e}{\cdot 10^} 
\newcommand{\dd}{\mbox{d}}

Transport networks, whether engineered by man or made by nature, need to maximize efficiency subject to cost constraints if they are to function optimally. Man-made distribution networks, such as those used in electric power distribution, irrigation or information flow can be centrally designed and constructed to achieve this optimality. However, in nature -- and occasionally in artificial networks -- the architecture of an efficient network is frequently the result of gradual alterations according to locally available information, rather than the execution of a central plan \cite{Gross2009, Rodriguez2001}. 

An example of such a system is the intracellular network of protoplasmic veins composing the slime mold \textit{Physarum polycephalum}. These veins perform cytoplasmic streaming and transport nutrients from food sources to the rest of the network. The organism redistributes the vein wall material so that little used conduits gradually disappear and heavily used ones grow. The physarum network exhibits a dense array of structurally organized loops and has been shown to be robust and efficient \cite{Tero2008}.

To address locally adaptive network systems, we adapted a mathematical model used for \textit{Physarum} in Refs.~\cite{Tero2008, Tero2006,Tero2007,Tero2010,Watanabe2011}. Like the \textit{Physarum} model, our model uses an undirected, weighted graph representation of the transport network, where the graph's edges represent the tubes and its nodes their junctions. The transport flows through the tubes are Hagen-Poiseuille and the tube conductivities dynamically adapt to the local flow. Note that only the edge weights are dynamically adapting, and not the nodes or the edges themselves. The underlying topology of the transport network thus remains the same unless edges disappear as their conductance approaches zero. Despite taking the \textit{Physarum} slime mold as our inspiration, however, we intend our analysis to be a more general exploration of possible network dynamics than an explicit model of \textit{Physarum polycephalum}.

In our analysis, the networks evolve from a number of prescribed initial network topologies and instead of spatially fixed sources, each pair of vertices of the network can act as a source and sink. The conductivity of the links grows or shrinks according to the average flow through them. As described in the rest of the text, starting from a random assignment of edge conductivities we find that the networks converge to a hierarchically organized architecture, dominated by a high conductivity backbone that spans the network. 
Additionally, the dynamical rearrangements of network links show striking, system-spanning behavior at certain key moments of the evolution that is reminiscent of glassy systems.

Significant variants of this basic adaptive networks concept can be attained in the formulation of the flow boundary conditions and in the details of the flow-based feedback for the conductivities. Both processes interact with one other -- the conductivity of an edge determines the flow through it, but the flow in turn gives rise to changes in the conductivity. Note that despite the adaptation of a given edge's conductivity being driven by the local flow conditions these conditions arise from the global solution of currents and pressure differentials over the entire network and hence some measure of global information affects the adaptation dynamics as well. Because of this interdependence it will be important to understand the type of interaction that exists between the conductivities and the flows, and in particular the associated timescales. It is known that \textit{Physarum}'s adaptation processes act on much longer timescales than its transportation processes \cite{Tero2008,Nakagaki2008}. Accordingly, we assume for our model that both the transport flows and the sources and sinks change much faster than the conductivities. The local flows are calculated as an ensemble average over the flows in a connected network of $N$ vertices created by each $N \cdot (N-1)/2$ possible pair of sources and sinks (cf.~Complete Multipoint Selection Method in \cite{Watanabe2011}). 

Our model is governed by a system of three types of equations, two for each edge -- describing transport and adaptation, respectively -- and one for each node, ensuring flow conservation. The flow $Q_{ij}^{kl}$ through the edge $\{i,j\}$ between the nodes $i$ and $j$ is the transported volume through the edge from node $i$ to node $j$ per unit time, when node $k$ is a source and node $l$ a sink. This quantity is also known as the volumetric flow rate or volume velocity, although $Q_{ij}^{kl}$ could just as well represent the amount of any transported good through a given route per time, like traffic on streets, data through a wireless connection, or electrons through a conductor. In \textit{Physarum}, the hydrostatic pressure difference $\Delta p_{ij}^{kl} := p_i^{kl}-p_j^{kl}$ along the tube between the pressures $p_i^{kl}$ and $p_j^{kl}$ defined at the nodes $i$ and $j$ acts as a potential difference from which the flow arises. The proportionality factor is the fraction of the tube's conductivity, denoted as $C_{ij}$, and its length $l_{ij}$, which gives the transport equation:
\begin{align}
    \label{eq:transportation}
    Q_{ij}^{kl} = \frac{C_{ij}}{l_{ij}} \cdot \left(p_i^{kl}-p_j^{kl}\right).
\end{align}
This first order flow can be thought of as a Hagen-Poiseuilleian laminar tube flow through a cylinder of length $l_{ij}$ and radius $r_{ij}$, for which the conductivity in the Newtonian case would be defined as $C_{ij} = \frac{\pi}{8 \eta} \cdot r_{ij}^4$, where $\eta$ is the dynamic viscosity of the fluid. The ensemble averaged mean flow can hence be calculated as
\begin{align}
    \label{eq:meanflow}
    \langle|Q_{ij}|\rangle := \frac{1}{\frac{N\cdot(N-1)}{2}} \sum_{(k,l) \in \mathbb{P}} \left|Q_{ij}^{kl}\right|,
\end{align} 
where $\mathbb{P}$ is the set of all node pairs, and flows are considered equally in both directions.

Meanwhile, for each node the sum of incoming flows must equal the sum of outgoing flows, unless the node is a source or sink which contributes an additional flow, $\zeta \ge 0$, defining the boundary conditions to eq. \ref{eq:transportation}:
\begin{align}
    \nonumber
    \sum_{j,  \forall \ \{i,j\} \in \mathbb{E}}  Q_{ij}^{kl} &= \left\{
                                \begin{array}{rl}
                                    \zeta & : i = k \\
                                    - \zeta & : i = l \\
                                    0 & : \mbox{else}
                                \end{array}
                            \right. \\ 
     \label{eq:conservation}        
                        &= (\delta_{ik}-\delta_{il}) \cdot \zeta.
\end{align} 
where $\mathbb{E}$ is the set of all edges.

Finally, we model the adaptation process with a differential equation describing the time evolution of the conductivities $C_{ij}=C_{ij}(t)$:
\begin{align}
    \label{eq:adaptation}    
    \frac{\dd C_{ij}(t)}{\dd t} = \beta \cdot f\left(\frac{\langle|Q_{ij}(t)|\rangle}{\epsilon}\right) - \alpha \cdot C_{ij}(t).
\end{align}
This equation features a positive, non-linear feedback term $\beta \cdot  f(\langle|Q_{ij}(t)|\rangle/\epsilon)$, that grows an edge's conductivity as a function of the scaled mean flow $\langle|Q_{ij}(t)|\rangle/\epsilon$ through itself. Balancing this term is a negative, exponential decay term $-\alpha \cdot C_{ij}(t)$. The parameters $\beta \ge 0$ and $\epsilon > 0$ scale the feedback and the flow through one edge; $\alpha \ge 0$ is the exponential decay parameter. We have chosen here not to co-evolve $l_{ij}$, though the manner in which changing tube lengths may embed into space is an interesting problem for future consideration.

The exact form of the feedback function $f$ must be chosen according to the details of the particular system under consideration. Nevertheless, it may be generalized to a function of $x:=\langle|Q_{ij}(t)|\rangle/\epsilon \ge 0$ that is positive definite, monotonically increasing, and bounded. Such a general $f$ would thus display no feedback when there is no flow, higher feedback for greater flows, and a capped maximum feedback, guaranteeing that $C_{ij}(t)$ does not diverge. These general conditions are fulfilled by a sigmoidal function, $f(x) = \frac{x^{\gamma}}{1+x^{\gamma}}$, which we use as our feedback function in this letter.

By appropriate non-dimensionalization we may reduce the control parameters for our dynamical system to $\vartheta := \zeta/\epsilon$, the load on the system, and $\gamma$, the sigmoidal exponent. Given this simplification, there are four possible ways to tune the system: different (random) initial conductivities $C_{ij}(0)$, different underlying initial topologies, varied transportation load $\vartheta$, or differently tuned sigmoidal feedback functions (varied $\gamma$). We have separately studied the influence of $\vartheta$ and $\gamma$. In all cases, multiple instances of different underlying topology classes -- Erd\H{o}s-Renyi (ER) pure random \cite{Erdos1959}, Watts-Strogatz (WS) small world \cite{Watts1998}, Barabasi-Albert (BA) rich-get-richer \cite{Barabasi1999} and 4-regular graph -- were examined, each with various initial value sets $C_{ij}(0)$.


\begin{figure}
\includegraphics[width=0.75\linewidth]{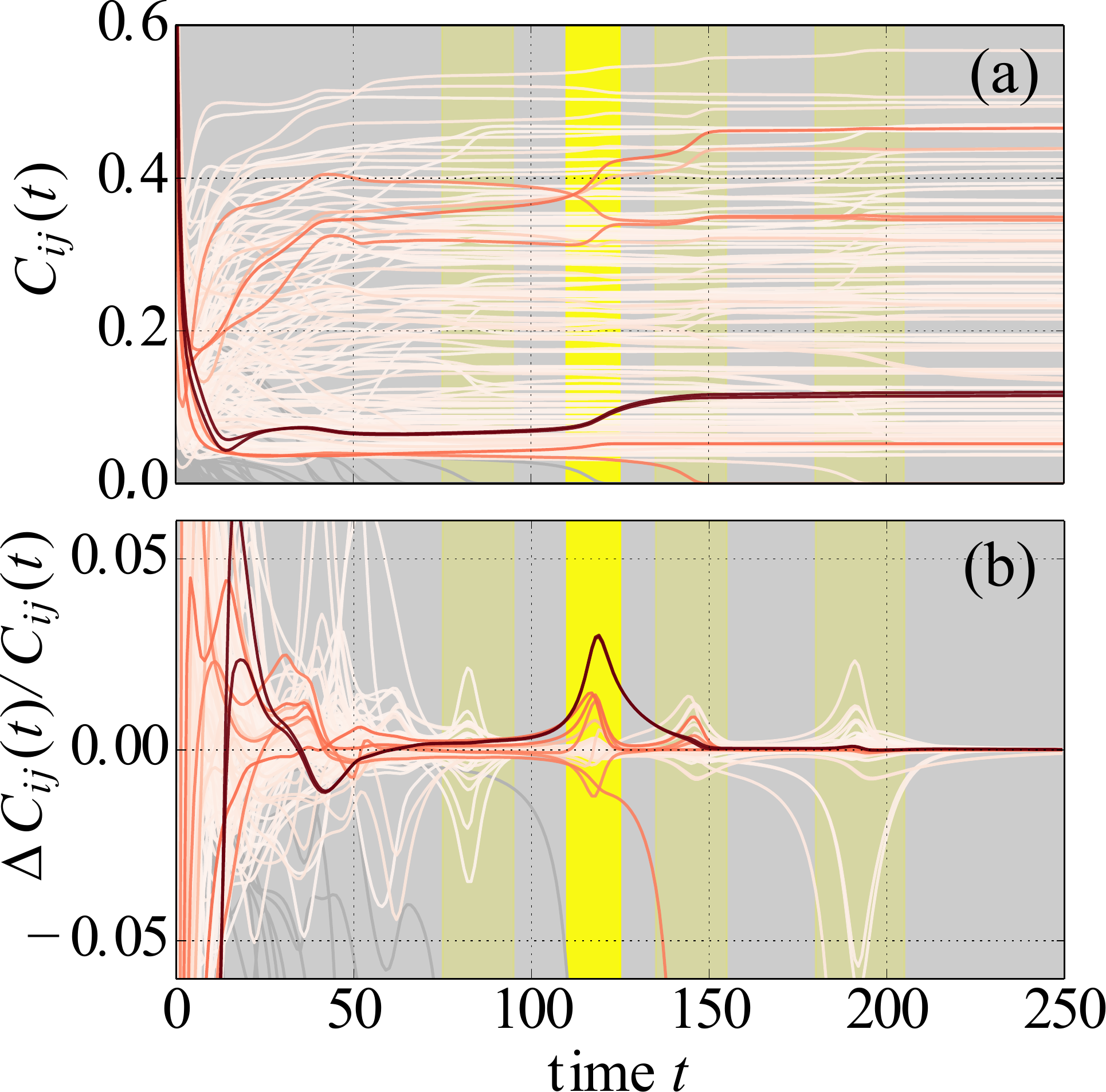}
\caption{\label{fig:conduct_vs_time} (Color online) (a) Time-evolution of the network conductivities. The conductivities are coloured based on the magnitude of the normalized maximum derivative of the conductivity (plotted in (b)) during the rearrangement event around $t=120$. After an initial period of transient dynamics, the system settles into a pattern of long times of slow or steady dynamics punctuated by rapid, avalanche-like rearrangements of the edge conductivities, such as the periods highlighted (in yellow). Only a subset of the edges in the network typically participate in each of these avalanche events.}
\end{figure}

After an initial rapid adaptation period, a typical simulation shows long periods of quiescent dynamics with little to no change in the conductivities, $C_{ij}(t)$, punctuated by rapid avalanche-like rearrangement events. In Fig. \ref{fig:conduct_vs_time} we plot a representative example of an ER network with $N=100$ nodes, load $\theta=10$ and feedback parameter $\gamma=1.8$. The slow equilibration dynamics is interrupted by short periods of fast adaptation, i.e. the period around $t=85, 120, 145$ and $190$ highlighted with yellow bands. This behavior is robust  and occurs regardless of the underlying topology classes or initial conductance values.

Note that once the conductivity of an edge drops below a certain threshold during the slow phase of the dynamics, it seems to enter a self-reinforcing runaway to null conductivity. As its flow continues to decrease the feedback term dies away and the conductivity can only decay exponentially thereafter. These feedback threshold events seem to trigger the rapid, short-timescale rearrangements, and further act as a selection effect on the edges, leading to a clean separation into two categories: regular, flow-conducting edges; and edges with insignificant flows and conductivities orders of magnitude smaller. We will refer to the former as `used edges' and the latter as `unused edges'. The event that an edge changes from being used to being unused can seemingly happen at any time, even after a long period of slow dynamics.

These repeatedly occurring selection and rearrangement processes disturb the dynamical system during its equilibration and change the final adapted state of the network. In general, the occurrence of these perturbations may prevent the system from reaching a final ground state in finite time, as can be the case in jammed or glassy dynamics. This kind of final state has been called a `quasi-attractor' or, in the context of adaptive networks, a `quasi-optimal solution' elsewhere \cite{Fricker2009, Gross2009}.

Meanwhile, the exponent $\gamma$ of the sigmoidal feedback function also acts on the dynamics of the conductivities independent of the underlying topology class. A constant feedback function ($\gamma=0$) uncouples the system of equations, leading to the same solution form $C_{ij}(t)$ for every edge in the network, independent of the flows:
\begin{align*}
	C_{ij}(t) = \frac{1}{2} + \left[ C_{ij}(0) - \frac{1}{2} \right] \cdot e^{-t} \quad \forall \ \{i,j\} \in \mathbb{E}. 
\end{align*}
Clearly, $\quad \lim_{t \to \infty} C_{ij}(t) = \frac{1}{2} \quad$ and hence all edges are used the same amount with no separation events. With increasing $\gamma$ the feedback becomes flow dependent and the coupling between the transport and the adaptation rises, resulting in full dynamics with increasing conductivity separation and initial value dependency (Fig. \ref{fig:conduct_vs_time}).
Interestingly however, this behavior is not observed for all non-zero values of $\gamma$. Instead, all edges are still in use for $0<\gamma<1$ and the final conductivities are not dependent on the initial values, but only determined by the edge's topological position in the network. All random initial value sets lead to the same final state. On the other hand, for $\gamma>1$ edges can become unused and the underlying topology of the adapting network effectively changes each time this occurs. As a consequence, the final state changes and the final conductivities are sensitive to the initial conditions in a non-linear, chaotic sense.

Higher exponents $\gamma \gg 1 $ sharpen the feedback function and the conductivity values separate further culminating in a complete, binary separation for $\gamma \to \infty$ where
\begin{align*}
	C_{ij}(t) = 1 + \left[ C_{ij}(0) -1 \right] \cdot e^{-t},
\end{align*}
for edges that remain used at all times and thus $ \lim_{t \to \infty} C_{ij}(t) = 1$ for this class of edges. However, these edges are not separable by an initial conductivity threshold value as might be expected. Instead, the initial value interval of the edges in use throughout overlaps that of those that become unused. It is therefore likely that the topological role of an edge is a primary contributor to its final use status.

\begin{figure}
\includegraphics[width=0.80\linewidth]{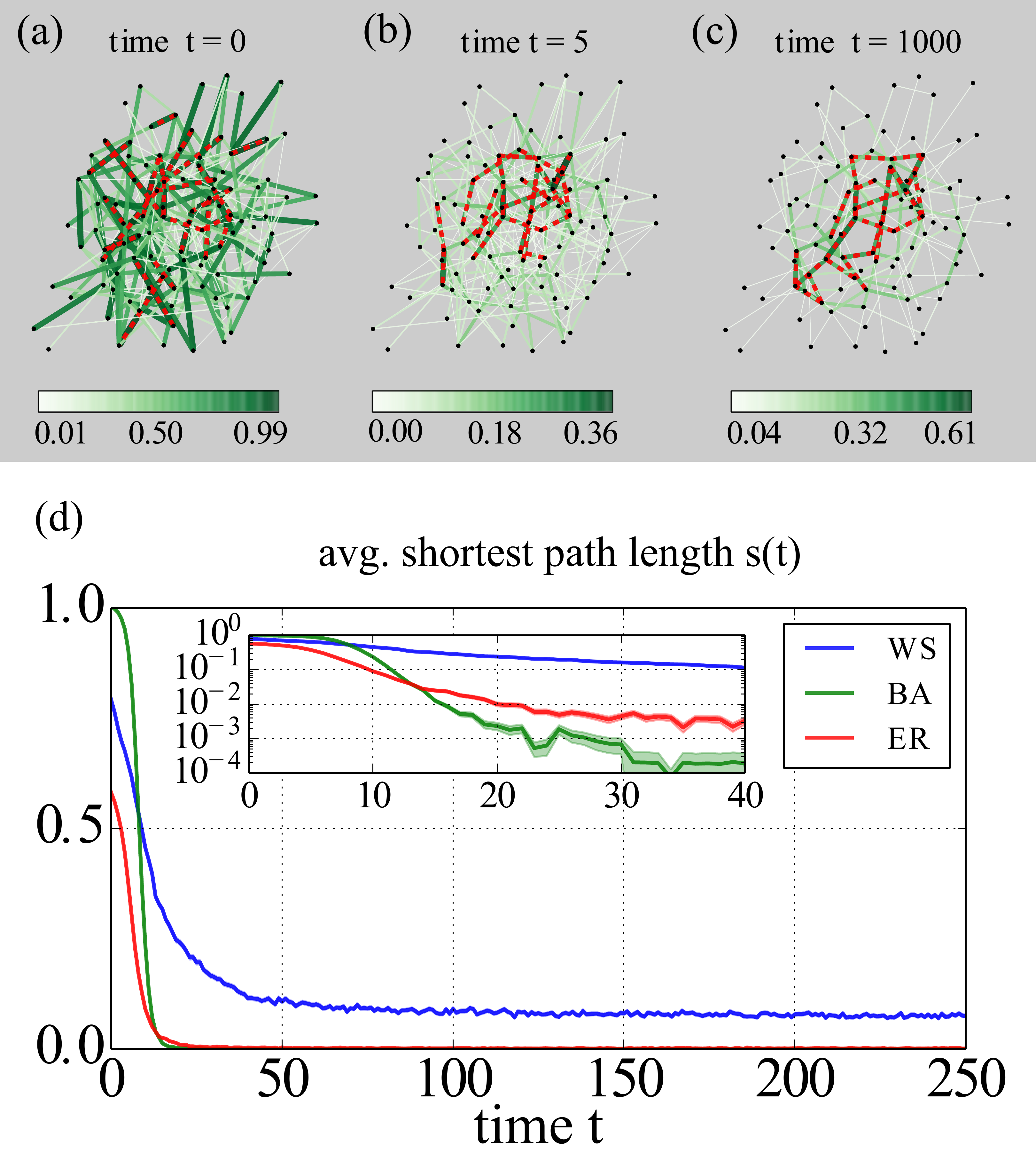}
\caption{\label{fig:backbone} (Color online) The development of a self-assembled backbone in the network conductivities under the action of the sigmoidal local feedback function. (a) The initial state of an ER network, with the top 10~\% conductive edges shown in red. These edges are randomly placed around the network with no additional underlying structure or connectivity. (b) The same network during the initial transient dynamics. Some rearrangement has occurred among the most conductive edges but no clear structure has yet emerged. (c) The network final state, after avalanche events have apparently ceased. The most conductive edges now seem to form a contiguous chain. (d) Normalized average shortest path length $s(t):= S(t)\cdot\langle 1/ S_{\mathrm{rand}}(t) \rangle$ for WS, BA and ER graphs. $S(t)$ is the average shortest path length of the graph with edge weights $1/C_{i,j}(t)$ and $\langle 1/S_{\mathrm{rand}}(t) \rangle$ is the mean inverse average shortest path length over 400 graphs generated by randomizations of the graph weights. $s(t)$ decreases as the backbone is formed.}
\end{figure}

%


To understand the influence of the adapting conductivities on the network architecture, it will thus be useful to relate them to their topological role. For that purpose, Fig.~\ref{fig:backbone} shows visualizations of an ER adapting network at different times in the evolution of the network dynamics. The plots were obtained by applying a spring layout algorithm \cite{Fruchterman1991, Hagberg2008} to the adapted state, for node positioning (black dots). The red dashed lines highlight the top 10~\% of all edges in the network.  In the initial transient phase the most conductive edges are disconnected edges at random positions. In the following adaptive phase, these edges get rearranged to different topological positions and create connected transportation backbones in the equilibration phase, indicating self-organized formation of hierarchical structure in the adapting network. Again, this behavior appears to be independent of the underlying topology class and the initial conductivity values (see Suppl. Material). 
To demonstrate more clearly the topological role of the backbone, in Fig.~\ref{fig:backbone}(d) we plot the normalised average shortest path length for WS, BA and ER adapting networks. The normalisation is necessary to decouple the effect of the adaptation of the overall conductivity magnitudes on the transportation efficiency of the network from the backbone creation. We observe that in all the topologies we considered $s(t)$ decreases very fast as a path minimising backbone is formed.

\begin{figure}
\includegraphics[width=0.80\linewidth]{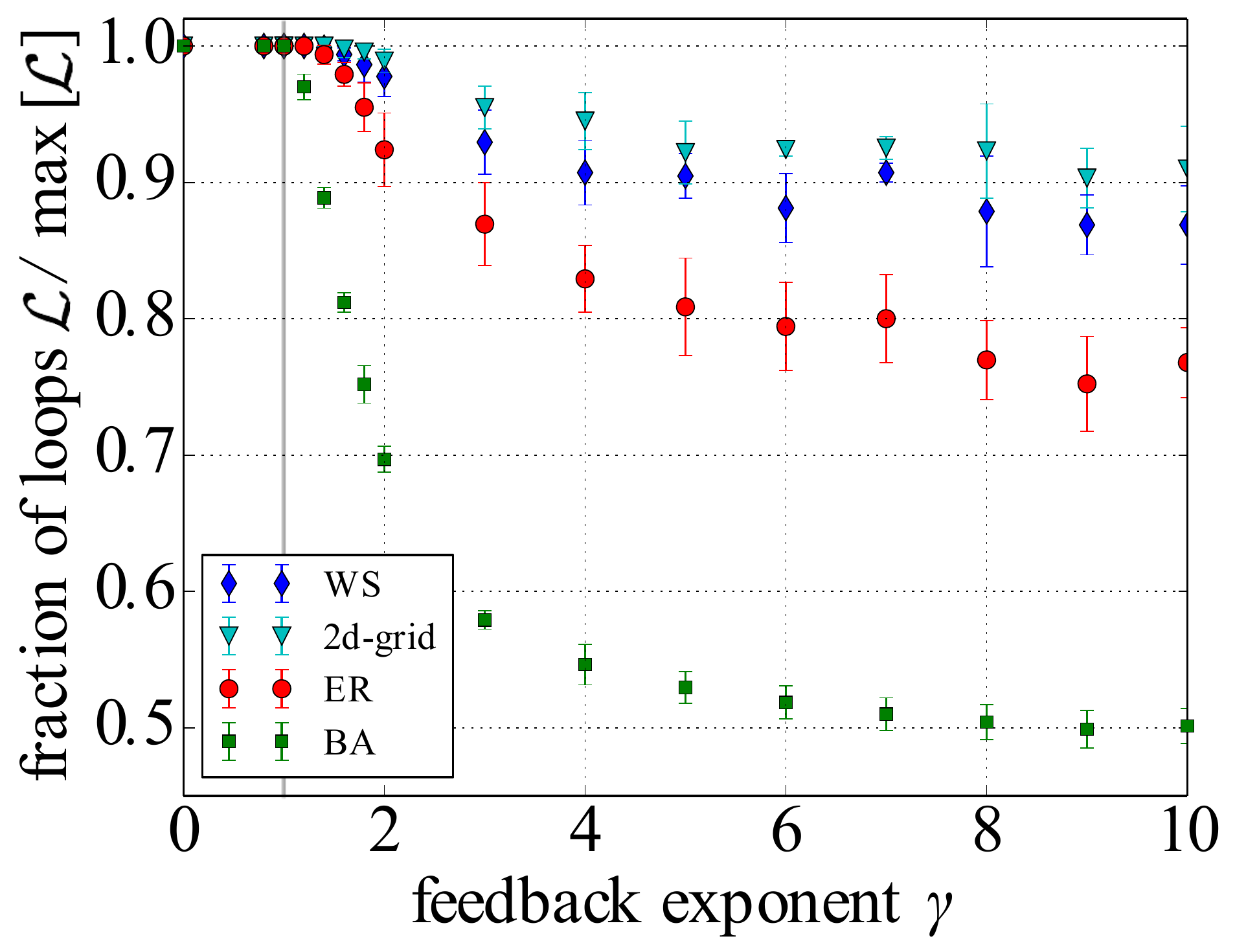}
\caption{\label{fig:loops} (Color online) The dependence of the evolved 'loopiness' of a network on the exponent, $\gamma$, in the sigmoidal feedback function. ER, WS, BA, and 4-regular initial underlying network topologies are all considered. In all cases there is a threshold value, $\gamma^* \approx 1$ (marked with a gray line) below which loop evolution does not occur. Above $\gamma^*$ all networks rapidly shed loops until $\mathcal{L}/\max(\mathcal{L})$ approaches a constant value $<1$, and significantly $<1$ in the case of the BA topology.}
\end{figure}

In addition to its role in the temporal conductivity dynamics, $\gamma$ also affects the topological character of the evolving network.
Keeping the number of nodes in the network constant, the number of fundamental loops $\mathcal{L}$ \cite{Veblen1912} changes with $\gamma$ as well, if only used edges, $E$, are considered, since $\mathcal{L} \sim E$. This allows a further characterization of the feedback threshold effect. 
Figure \ref{fig:loops} shows the mean and the standard deviation values of $\mathcal{L}(\gamma)$ from all performed simulations. Several underlying topologies for each class and multiple initial conductivity value sets are included. To compare across different topology classes the values have been normalized by the according maximum number of loops under the assumption that every existing edge is used. The relatively sharp threshold at $\gamma^{*} \approx 1$ appears to describe a transition point between a phase with $\mathcal{L}=\max[\mathcal{L}]$, independent of the underlying topology, for $\gamma < \gamma^{*}$ and a topology dependent phase for $\gamma>\gamma^{*}$. The BA networks show a sharper transition with significantly lower $\mathcal{L}$ values for $\gamma>\gamma^{*}$ than the rest, using only half of the available loops for the adapted network architecture in the limit where the feedback function is step-like: $\mathcal{L}(\gamma \to \infty)\approx0.5\cdot\max[\mathcal{L}]$.
To compare the randomly generated network topologies (BA, ER, WS) with a completely regular structure, Fig.~\ref{fig:loops} also includes data simulated on a two-dimensional grid with periodic boundary conditions (4-regular grid). The regular grid produced an adapted state with very few unused edges, expressed in a high number of loops $\mathcal{L}(\gamma \to \infty)\approx0.9\cdot\max[\mathcal{L}]$. It is remarkable then that the topologies with high clustering coefficients and average degrees -- as well as low average path lengths -- also show a higher number of final-state loops and vice versa, including the regular grid.

We have demonstrated that a transport network evolving under the influence of local information can and generically does exhibit very rich behavior, from the emergent self-assembly of a conductivity backbone with topology-dependent hierarchical scaffolding in the network's architecture to a phase transition-like sensitivity to the control exponent in the feedback function and glassy dynamics complete with separation of timescales and avalanches of conductivity rearrangements. Despite the local character of the adaptation with respect to the conductivities, the flow that feeds this adaptation carries some measure of global information, and it is this global information that in principle permits a scale-free hierarchical architecture instead of the length-scale restricted features one may have expected from purely local reaction-diffusion pattern formation. Ultimately, these features together with the dynamical richness in the system may help provide insight not only into the adaptive network of the \textit{Physarum polycephalum} slime mold that inspired this study, but also to other adapting or developing distribution networks that leverage this kind of `local-yet-global' information such as mesoscale angiogenesis or social networks. 

Despite our choice of a Hagen-Poiseuille fluid flow model for our study, we believe the scope of the phenomenon described in this letter to be general. Therefore, with an aim to deepening our understanding of the universal features of self-organized adaptation in transport networks we have begun to expand our attention to other relevant models for distribution networks, such as a shortest-path usage model, and a capacity based Ford-Fulkerson maximum flow model \cite{Ford1956}. An understanding of the fundamental physical differences between local update rulesets and how those differences play out in a network's adaptation dynamics is an important first step in both describing these complex systems and in designing computational or communicational analogs.

\begin{acknowledgments}
CM and MM acknowledge support from the National Science Foundation under grant ID~1058899 and the Simons Foundation. EK wishes to thank the Burroughs Wellcome Fund.
\end{acknowledgments}


\bibliography{KatiforiAdaptNets} 

%
%
\end{document}